\documentclass[11pt,a4paper,twoside]{article}
\usepackage{axodraw}
\usepackage{color}
\usepackage[latin1]{inputenc}
\usepackage{amsfonts}
\usepackage{amsmath}
\usepackage{amssymb}
\usepackage{amsthm}
\usepackage{cite} 
\usepackage[dvips]{graphicx}
\usepackage{fleqn} 
\usepackage[footnotesize]{caption}
\usepackage{cancel}
\usepackage[center,footnotesize,hang]{subfigure}
\newcommand{\raw}{\rightarrow}

\begin{document}

\begin{titlepage}

\vspace*{-15mm}
\begin{flushright}
MPP-2008-74\\
\end{flushright}
\vspace*{0.7cm}

\begin{center}
{
\bf\LARGE Non-Standard Neutrino Interactions with Matter\\[2mm]
from Physics Beyond the Standard Model
}
\\[8mm]

S.~Antusch$^{\star}$
\footnote{E-mail: \texttt{antusch@mppmu.mpg.de}}, 
J.~P.~Baumann$^{\star\dagger} $
\footnote{E-mail: \texttt{jbaumann@mppmu.mpg.de}}, 
E.~Fern\'andez-Mart\'{\i}nez$^{\star}$
\footnote{E-mail: \texttt{enfmarti@mppmu.mpg.de}},
\\[1mm]

\end{center}
\vspace*{0.50cm}
\centerline{$^{\star}$ \it 
Max-Planck-Institut f\"ur Physik (Werner-Heisenberg-Institut),}
\centerline{\it 
F\"ohringer Ring 6, D-80805 M\"unchen, Germany}
\vspace*{0.1cm}
\centerline{$^\dagger$ \it Arnold Sommerfeld Center, Department f\"ur Physik,}
\centerline{\it 
Ludwig-Maximilians-Universit\"at M\"unchen,}
\centerline{\it 
Theresienstra\ss{}e 37, D-80333 M\"unchen, Germany}
\vspace*{1.20cm}

\begin{abstract}

\noindent 
We investigate how non-standard neutrino interactions (NSIs) with matter can be generated 
by new physics beyond the Standard Model (SM) and analyse the constraints on the NSIs in 
these SM extensions. We focus on tree-level realisations of lepton number conserving 
dimension 6 and 8 operators which do not induce new interactions of four charged fermions 
(since these are already quite constrained) and discard the possibility of cancellations 
between diagrams with different messenger particles to circumvent experimental constraints.  
The cases studied include classes of dimension 8 operators which are often referred to as 
examples for ways to generate large NSIs with matter. We find that, in the considered 
scenarios, the NSIs with matter are considerably more constrained than often assumed in 
phenomenological studies, at least ${\cal O}(10^{-2})$. The constraints on the 
flavour-conserving NSIs turn out to be even stronger than the ones for operators which 
also produce interactions of four charged fermions at the same level. Furthermore, we find 
that in all studied cases the generation of NSIs with matter also gives rise to NSIs at the 
source and/or detector of a possible future Neutrino Factory.    
\end{abstract}

\end{titlepage}

\newpage
\setcounter{footnote}{0}

\section{Introduction}
With the near start of the LHC, particle physics will enter a new era. With unprecedented energy reach and luminosity, the LHC will allow to clarify the origin of electroweak symmetry breaking and look for new physics at TeV energies. In addition, complementary to the LHC, future precision experiments in the neutrino sector are aiming at measuring the remaining unknown parameters in the lepton sector, i.e.\ the neutrino mass scale,  leptonic CP violation and the remaining unknown leptonic mixing angle $\theta_{13}$ \cite{Bandyopadhyay:2007kx}. 

Precision neutrino oscillation experiments, for example, are also sensitive to new physics beyond the Standard Model (SM). This sensitivity is at the same time a chance and a potential problem:
On the one hand, there is the chance that such experiments discover new physics, for example new interactions of neutrinos at the source, detector or with matter, a possible non-unitary leptonic mixing matrix, or even a violation of fundamental principles such as CPT invariance or locality.     
On the other hand, new physics may also lead to confusions of effects from new (CP violating) interactions with the leptonic Dirac CP phase, in the standard parameterisation of the leptonic mixing matrix, or with the small mixing parameter $\theta_{13}$. 
To avoid such confusion when measuring the remaining unknown parameters in the lepton sector, a better knowledge of the constraints on the new physics relevant to these experiments is highly desirable.  

With respect to their effects on neutrino oscillations, one convenient way to describe new interactions with neutrinos in the electroweak (EW) broken phase are the so-called NSI parameters for non-standard neutrino interactions at the source ($\varepsilon^s_{\alpha\beta}$), detector ($\varepsilon^d_{\alpha\beta}$) \cite{newint} and with matter ($\varepsilon^m_{\alpha\beta}$) \cite{Campanelli:2002cc,Davidson:2003ha,yasuda}. They give the relative strength of these interactions with respect to the Fermi constant $G_F$. Among these parameters, the NSIs with matter are comparatively weakly constrained, i.e.\ some bounds on them are even ${\cal O}(1)$ \cite{Davidson:2003ha}. In many analyses, large non-standard matter effects are therefore included, whereas possible new interactions at the source or detector are set to zero. 

In this study, we investigate how non-standard neutrino interactions (NSIs) with matter \cite{Wolfenstein:1977ue} can be induced by new physics beyond the SM. 
We restrict our analysis to tree-level realisations of lepton number conserving dimension 6 and 8 operators which do not induce new interactions of four charged fermions (since these are already quite constrained), and discard the possibility of cancellations between diagrams with different messenger particles to circumvent constraints.  
The cases studied include classes of dimension 8 operators which are often referred to as examples for ways to generate large NSIs with matter without generating NSIs at the source and detector of a neutrino oscillation experiment \cite{Davidson:2003ha,BerezhianiRossi01}.   
The goal of this study is to investigate the constraints on the NSI parameters if these operators are generated by explicit new physics beyond the SM.

The paper is organised as follows: In section 2 we define the NSIs $\varepsilon^{m,f}_{\alpha\beta}$ with matter and the related quantities $\tilde{\varepsilon}^{m}_{\alpha\beta}$ which affect neutrino oscillations in matter. 
In section 3 we discuss the possible approaches to realising large NSIs with matter as well as the restrictions we impose on our analysis. 
The generation of matter NSIs from dimension 6 operators is discussed in section 4, and updated and improved constraints on the NSI parameters are derived. 
In section 5 we investigate NSIs with matter from dimension 8 operators and derive the corresponding constraints. Section 6 contains a summary and our conclusions.   

\section{NSIs with matter}
Compared to the bounds on NSIs at the source and detector, the NSIs which can modify matter effects are often assumed to be only very weakly constrained. In the following we will therefore mainly restrict ourselves to this class of NSIs. 
The (lepton number conserving) NSI four-fermion operators of interest are contained in the following Lagrangian after EW symmetry breaking,
\begin{eqnarray}\label{Eq:NSIs_matter}
{\cal L}^m_{\mathrm{NSI}} &=& 2 \sqrt{2} G_F 
\sum_{f} \varepsilon^{m,f_\mathrm{L}}_{\alpha\beta} 
\left[ \bar \nu_{\mathrm{L}\alpha} \gamma^\delta \nu_{\mathrm{L}\beta} \right] 
\left[ \bar f_{\mathrm{L}} \gamma_\delta f_{\mathrm{L}}\right]  \nonumber \\
&&+
2 \sqrt{2} G_F \sum_{f} \varepsilon^{m,f_\mathrm{R}}_{\alpha\beta} 
\left[ \bar \nu_{\mathrm{L}\alpha} \gamma^\delta \nu_{\mathrm{L}\beta} \right] 
\left[ \bar f_{\mathrm{R}} \gamma_\delta f_{\mathrm{R}} \right] \!.
\end{eqnarray}
The fermions $f$ which the neutrinos couple to are either electrons $e$, up quarks $u$ or down quarks $d$, and may be left- or right-handed. $\alpha,\beta=1,2,3$ are family indices. 
Constraints on the parameters $\varepsilon^{m,f}_{\alpha\beta}$ have been derived in \cite{Davidson:2003ha,Barranco:2005ps}.

Neutrino oscillations in the presence of non-standard matter effects can be described by an effective square mass matrix which can be parameterised as
\begin{eqnarray}
M_{\mathrm{eff}}^2 = U_{\mathrm{PMNS}}\cdot \mbox{diag}(0,\Delta m^2_{21},\Delta m^2_{31})\cdot U_{\mathrm{PMNS}}^\dagger  + 2 E V (\mbox{diag}(1,0,0) + \tilde{\varepsilon}^m_{\alpha\beta}) \;,
\end{eqnarray}
where $V=\sqrt{2} G_F n_e$, with $n_e$ being the electron number density. The parameters $\tilde{\varepsilon}^m_{\alpha\beta}$ are given by
\begin{eqnarray}\label{Eq:DefEpsTilde}
\varepsilon^{m,f}_{\alpha\beta} = \varepsilon^{m,f_\mathrm{L}}_{\alpha\beta} + \varepsilon^{m,f_\mathrm{R}}_{\alpha\beta}\;,\quad
 \tilde{\varepsilon}^m_{\alpha\beta} = 
 \varepsilon^{m,e}_{\alpha\beta} + 2\varepsilon^{m,u}_{\alpha\beta}+ \varepsilon^{m,d}_{\alpha\beta}+
 \frac{n_n}{n_e}(\varepsilon^{m,u}_{\alpha\beta}+ 2\varepsilon^{m,d}_{\alpha\beta})\;,
\end{eqnarray}
where $n_n$ is the neutron number density. While the individual $\varepsilon^{m,f_\mathrm{L}}_{\alpha\beta}$ and $\varepsilon^{m,f_\mathrm{R}}_{\alpha\beta}$ are predicted in an explicit extension of the SM, only the combined quantity $\tilde{\varepsilon}^m_{\alpha\beta}$ is relevant for neutrino oscillations in matter.

\section{Strategy to realise large NSIs with matter}\label{sec:restrictions}

Direct experimental constraints on the effective operators of Eq.~(\ref{Eq:NSIs_matter}) can mainly be derived through neutrino scattering experiments off electrons or nuclei and are, therefore, rather weak. The main goal of this work is to investigate whether these mild bounds can be saturated in extensions of the Standard Model avoiding stronger constraints from other operators generated by the same SM extension as a byproduct. Indeed the operator of Eq.~(\ref{Eq:NSIs_matter}) is not gauge invariant and its generation in any extension of the SM will usually involve also the charged lepton partners of the neutrinos in the SU(2)$_\mathrm{L}$ doublets. The simplest possibility to generate NSIs with matter by SU(2)$_{\mathrm{L}}$-invariant operators would be 
to promote the left-chiral leptons (quarks) in the effective operators of Eq.~(\ref{Eq:NSIs_matter}) to lepton (quark) doublets.
However, these SU(2)$_{\mathrm{L}}$-invariant operators also generate interactions of four charged fermions. Conservatively estimated constraints on the relevant off-diagonal NSIs with matter from such operators range from ${\cal O}(10^{-2})$ ($\varepsilon^{m,u,d}_{e\tau},\varepsilon^{m,u,d}_{\mu\tau}$) to ${\cal O}(10^{-6})$ ($\varepsilon^{m,e}_{e\mu}$) for the off-diagonal elements \cite{Bergmann:1997mr} - \cite{Ibarra:2004pe}, while the constraints for the diagonal elements
are rather weak (but still stronger than the direct bounds of \cite{Davidson:2003ha,Barranco:2005ps}).\footnote{Further relaxation of these bounds (up to a factor of $7$) is in principle possible \cite{Bergmann:1999rz,Bergmann:2000gp}, however it would require specific arrangements in SU(2)$_{\mathrm{L}}$-breaking.} 
In the following, we will therefore restrict ourselves to extensions of the SM (i.e.\ to the introduction of additional messenger particles and interactions) where no interactions of four charged fermions are generated at tree-level. 
For example, this excludes SU(2)$_\mathrm{L}$-triplet fermions or scalars as messengers. Their low energy effects include interactions of four charged leptons and the resulting NSIs are known to be subject to the above constrains (see e.g.\ \cite{Abada:2007ux}). 

We will furthermore not consider here the possibility that extensions of the Standard Model with different messenger particles induce these constrained four charged fermion operators but conspire to cancel against each other circumventing the bounds. Such cancellations between contributions from different messenger particles with different SM quantum numbers are typically associated with a certain amount of fine-tuning without justification by a suitable symmetry. 
We will consequently disregard the extensions in which the extra particles and couplings introduced to produce the NSIs of Eq.~(\ref{Eq:NSIs_matter}) also lead to diagrams with four charged fermions. 
More generally, we do not consider the possibility that experimental constraints are circumvented by cancellations between contributions from different messenger particles.

The lowest dimensional gauge invariant realisations  
of the effective interactions of Eq.~(\ref{Eq:NSIs_matter}), and presumably the least suppressed ones, are  provided by dimension 6 operators.  Loop-level generation of them will always be suppressed by an additional ``loop factor'' of ${\cal O}(1/16\pi^2)$ and cannot lead to very large NSIs. We will therefore restrict ourselves to tree-level generation of effective operators. On the other hand, certain dimension 8 operators \cite{Davidson:2003ha,BerezhianiRossi01} containing four fermions and two SM Higgs fields are often quoted as examples how to generate large NSIs (less constrained than NSIs from dimension 6 operators). The reason why dimension 8 operators appear promising is that the vev of the Higgs SU(2)$_{\mathrm{L}}$ doublets can be used to ``project out'' the neutrino fields from the SU(2)$_{\mathrm{L}}$ doublets after electroweak symmetry breaking, thus avoiding to generate interactions with charged leptons instead of neutrinos. 
In our analysis, we therefore also include the class of dimension 8 operators (which contains the above-mentioned operators) with external fields $L,\bar L, f, \bar f, H, H^\dagger$, where $f$ can be $f_\mathrm{L}$ or $f^c_\mathrm{R}$  with $f_\mathrm{L} \in \{L_1,Q_1\}$ and $f^c_\mathrm{R} \in \{e^c_\mathrm{R},u^c_\mathrm{R},d^c_\mathrm{R}\}$ and where $H$ is the SM Higgs doublet. 

\vspace{0.2cm}

In summary, when we attempt to realise large NSIs with matter, we will restrict our search to extensions of the SM satisfying the following {\em restrictions}:

\begin{itemize} 

\item {\bf No new interactions of four charged fermions}

\item {\bf No cancellations between diagrams with different messenger particles}

\item {\bf Tree-level generation of the NSIs through dimension 6 and 8 operators}

\item {\bf Electroweak symmetry breaking is realised via the Higgs mechanism}

\end{itemize} 

In the following, we will always impose these restrictions on our analysis. 
We will start with dimension 6 operators and their generation in section \ref{Sec:dim6} and turn to the class of dimension 8 operators in section \ref{Sec:Dim8}. 

\section{Dimension 6 operator for matter NSIs}\label{Sec:dim6}

There are only two dimension 6 operators and associated SM extensions satisfying the criteria defined in the previous section: 
the anti-symmetric 4-lepton operator, generated from the exchange of virtual singly charged scalar fields (c.f.\ figure \ref{fig:d=6antisymm}), and  the dimension 6 operator modifying the neutrino kinetic terms, generated by the exchange of virtual fermionic singlets (c.f.\ figure \ref{fig:d=6kin}). The latter operator generates the NSIs in an indirect way, i.e.\ after canonical normalisation of the neutrino kinetic terms.

\subsection{Constraints on the NSIs from the anti-symmetric dimension 6 operator}\label{App:AnitSymmDim6}
\begin{figure}
   \centering
   \includegraphics[scale=0.7]{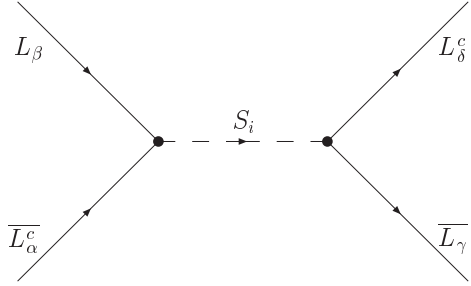}
   \caption{
   Generation of the anti-symmetric 4-lepton operator by the exchange of virtual singly charged scalars $S_i$.
   \label{fig:d=6antisymm}}   
\end{figure}

In this subsection we will review and update the bounds on the matter NSIs generated from the anti-symmetric dimension 6 operator composed of four lepton doublets (see also: \cite{Bilenky:1993bt,Bergmann:1999pk,Cuypers:1996ia})
\begin{eqnarray}\label{Eq:AntisymmDim6}
{\cal L}^{d=6,as}_{NSI} =
c^{d=6,as}_{\alpha\beta\gamma\delta} (\overline{L^c}_\alpha \cdot L_\beta) (\bar L_\gamma \cdot L^c_\delta) 
\;,
\end{eqnarray} 
considering its tree-level generation via singly charged scalar fields $S_i$ (c.f.\ figure \ref{fig:d=6antisymm}), i.e.\ new fields beyond the SM in the representation $(1,1,-1)$ of the SM gauge group $G_{321}=$ SU(3)$_{\mathrm{C}}\times$SU(2)$_{\mathrm{L}}\times$U(1)$_{\mathrm{Y}}$. 
The dot in Eq.~(\ref{Eq:AntisymmDim6}) denotes the SU(2)$_\mathrm{L}$ invariant product (where indices are contracted with $\varepsilon := i \sigma_2$). 
In addition to the SM Lagrangian, we therefore consider the additional interaction
\begin{equation}\label{Eq:IntS}
 \mathcal{L}^{ S }_{int} = - \lambda^i_{\alpha\beta} \overline{L}_{\alpha}^c i \sigma_2 L_\beta  S_i  + \text{H.c.} = \lambda^i_{\alpha\beta}  S_i 
(\overline{\ell}^c_\alpha P_L \nu_\beta - \overline{\ell}^c_\beta P_L \nu_\alpha) + \text{H.c.}
\label{antisd6}
\end{equation}
as well as a mass $m_{S_i}$ for the $S_i$.

Integrating out the heavy scalars $S_i$ generates the dimension 6 operator of Eq.~(\ref{Eq:AntisymmDim6}) at tree level. Written in component fields, it has the form
\begin{eqnarray}
{\cal L}^{d=6,as}_{NSI} =
4\sum_i\frac{\lambda^i_{\alpha \beta} \lambda^{i*}_{\delta \gamma}}{m_{S_i}^2} (\overline{\ell^c}_\alpha P_L \nu_\beta) (\bar \nu_\gamma P_R \ell^c_\delta) =  
2\sum_i\frac{\lambda^i_{\alpha \beta} \lambda^{i*}_{\delta \gamma}}{m_{S_i}^2} (\overline{\ell}_\delta \gamma^\mu P_L \ell_\alpha) (\bar \nu_\gamma \gamma_\mu P_L \nu_\beta)
\;.
\label{antisd6comp}
\end{eqnarray}
For the coefficients $c^{d=6,as}_{\alpha\beta\gamma\delta}$ we can read off
\begin{eqnarray}
c^{d=6,as}_{\alpha\beta\gamma\delta} = - \sum_i \frac{\lambda^i_{\alpha \beta} \lambda^{i*}_{\delta \gamma}}{m_{S_i}^2}
\;.
\end{eqnarray}
Using the definition of Eq.~(\ref{Eq:NSIs_matter}), we find that, for normal matter, only the NSIs 
\begin{eqnarray}
\varepsilon^{m,e_\mathrm{L}}_{\alpha \beta}  =  
\sum_i \frac{\lambda^i_{e \beta} \lambda^{i*}_{e \alpha}}{\sqrt{2} G_F m_{S_i}^2} 
\label{epsilond6}
\end{eqnarray}
are induced. 
We note that, since the coupling matrix $\lambda^i_{\alpha \beta}$ is anti-symmetric, the indices $\alpha$ and $\beta$ in $\varepsilon^{m,e_\mathrm{L}}_{\alpha \beta}$ satisfy $\alpha, \beta \neq e$.\footnote{Analogously, the operator in Eq.~(\ref{Eq:AntisymmDim6}) also induces the NSIs $\varepsilon^{m,\mu_\mathrm{L}}_{\alpha \beta}$ ($\alpha, \beta \neq \mu$) and $\varepsilon^{m,\tau_\mathrm{L}}_{\alpha \beta}$ ($\alpha, \beta \neq \tau$), which do not play a role for neutrino oscillations in normal matter. The generalisation of the constraints to these NSIs is straightforward.}

\subsubsection{Bounds from rare lepton decays}
One type of constraints in the above extension of the SM comes from rare radiative lepton decays $l_\alpha \longrightarrow l_\beta \gamma $. Neglecting the masses of the light leptons we obtain
\begin{equation}
 \frac{\Gamma(l_\alpha \raw l_\beta \gamma)}{\Gamma(l_\alpha \raw l_\beta \nu_\alpha \bar{\nu}_\beta)} = \frac{\alpha}{48 \pi} \left| \sum_i \frac{\lambda^i_{\alpha \delta} \lambda^{i*}_{\beta \delta}}{m_{S_i}^2 G_F} \right|^2\;,
\end{equation}
with $\delta \neq \alpha,\beta$. Using the present experimental
bounds \cite{muegamma} at 90\% confidence level (cl)
\begin{eqnarray}
Br (\mu  \rightarrow e \gamma) &<& 1.2 \cdot 10^{-11}
\; ,\\
Br (\tau \rightarrow e \gamma) &<& 9.4 \cdot 10^{-8}
\; ,\\
Br (\tau \rightarrow \mu \gamma) &<& 1.6 \cdot 10^{-8}
\; ,
\end{eqnarray}
together with the experimental values $Br (\tau \rightarrow \nu_\tau \mu
\overline{\nu}_\mu) = 0.1736 \pm
0.0006$, $Br (\tau \rightarrow \nu_\tau e \overline{\nu}_e) = 0.1784 \pm 0.0006$ and $Br (\mu
\rightarrow \nu_\mu e \overline{\nu}_e) \approx 100$\%~\cite{PDG}, we obtain the following constraints:
\begin{eqnarray}\label{raredecaysd6x}
\left| \sum_i \frac{\lambda^i_{e \tau} \lambda^{i*}_{\mu \tau}}{m_{S_i}^2 G_F} \right| &<& 5.0 \cdot 10^{-4}
\; ,\\
\left| \sum_i \frac{\lambda^i_{e \mu} \lambda^{i*}_{\mu \tau}}{m_{S_i}^2 G_F} \right| &<& 1.0 \cdot 10^{-1}
\; ,\\
\left| \sum_i \frac{\lambda^i_{e \mu} \lambda^{i*}_{e \tau}}{m_{S_i}^2 G_F} \right| &<& 4.4 \cdot 10^{-2}
\; .
\label{raredecaysd6}
\end{eqnarray}
Comparing them with Eq.~(\ref{epsilond6}) we see that only $\tau \raw \mu \gamma$ allows to constrain one of the matter NSI parameters. At the 90 \% cl this constraint is given by
\begin{eqnarray}
|\varepsilon^{m,e_\mathrm{L}}_{\mu \tau}| &<& 3.0 \cdot 10^{-2}\;.
\end{eqnarray}
This bound turns out to be comparatively weak compared to the bounds that can be obtained from the  determination of $G_F$ via $\mu$ and $\tau$ decays under the assumption of unitarity of the CKM matrix, as we will now discuss. 

\subsubsection{Bounds from $G_F$ via $\mu$ and $\tau$ decays and assuming CKM unitarity}
\label{CKM}

The unitarity constraint on the first row of the CKM matrix is experimentally tested to very high precision. The extraction of $V_{ud}$ is performed through superallowed $\beta$ decays, while $V_{us}$ is measured through kaon decays.\footnote{The experimental value of $V_{ub}$ is smaller than the precision of the other two matrix elements in the unitarity relation and thus negligible for this discussion.} 
In both processes $G_F$, extracted from $\mu$ decays, is used as an input. Thus, if we assume that the CKM matrix is unitary, the experimental bounds provide excellent constraints on new physics contributions to $\mu$ decays.

The singly charged scalars $S_i$ introduced in Eq.~(\ref{antisd6}) can mediate the decay $\mu \raw e \nu_\alpha \bar{\nu}_\beta$ with $\alpha \neq e$ and $\beta \neq \mu$. For $\alpha = \mu$ and $\beta = e$, in particular, the diagram interferes with the SM decay amplitude and the suppression of the process will be linear in each $\lambda^i_{\alpha \beta}$ instead of quadratic. At this order in $\lambda^i_{\alpha \beta}$, the Fermi constant extracted from the $\mu$ decay would be given by
\begin{equation}
 G_\mu = G_F(1 + \sum_i \frac{|\lambda^i_{e \mu}|^2}{\sqrt{2}m_{S_i}^2 G_F}) = G_F(1 + \varepsilon^{m,e_\mathrm{L}}_{\mu \mu})\;.
\end{equation}
Using $G_\mu$ to extract the values of $V_{ud}$ and $V_{us}$ from $\beta$ decays and kaon decays leads to
\begin{equation}
V^{exp}_{\alpha \beta} = \frac{V_{\alpha \beta}}{1 + \varepsilon^{m,e_\mathrm{L}}_{\mu \mu}}\;,
\end{equation}
where $V^{exp}_{\alpha \beta}$ denotes the experimentally measured $V_{ud}$ and $V_{us}$.
Using \cite{PDG}
\begin{eqnarray}
 V_{ud}^{exp} &=& 0.97418 \pm 0.00027\;, \\ 
 V_{us}^{exp} &=& 0.2255 \pm 0.0019\;, 
\end{eqnarray}
and assuming that the unitarity of the CKM matrix is not affected by the new physics leading to the NSIs, we find
\begin{equation}
|V_{ud}^{exp}|^2 + |V_{us}^{exp}|^2  = 
\frac{1}{\left( 1 + \varepsilon^{m,e_\mathrm{L}}_{\mu \mu} \right) ^2}  = 0.9997 \pm 0.0010\;.
\label{constr1}
\end{equation}
Analogous to the case of the $\mu$ decay, the decay $\tau \raw e \nu \bar\nu$ is modified to 
\begin{equation}
 G_{\tau \rightarrow e \nu \bar\nu} = G_F (1 + \sum_i \frac{|\lambda^i_{e \tau}|^2}{\sqrt{2}m_{S_i}^2 G_F}) = G_F(1 + \varepsilon^{m,e_\mathrm{L}}_{\tau \tau})\;.
\end{equation}
The comparison with the $\mu$ decay can now be used to obtain bounds on the universality of the weak interactions \cite{PDG,universality}, which yields
\begin{equation}
\sqrt{\frac{G_{\tau \rightarrow e \nu \bar\nu}}{G_{\mu \rightarrow e \nu 
\bar\nu}}} = \sqrt{\frac{1 + \varepsilon^{m,e_\mathrm{L}}_{\tau \tau}}{1 + \varepsilon^{m,e_\mathrm{L}}_{\mu \mu}}}=1.0004 \pm 0.0023\;.
\label{constr2}
\end{equation}

\subsubsection{Constraints on NSIs with matter}
Using Eqs.~(\ref{constr1}) and (\ref{constr2}) and additionally the relation $|\varepsilon^{m,e_\mathrm{L}}_{\mu \tau}| \leq \sqrt{\varepsilon^{m,e_\mathrm{L}}_{\mu \mu} \varepsilon^{m,e_\mathrm{L}}_{\tau \tau}}$ derived from Eq.~(\ref{epsilond6}), 
we obtain the following bounds (at 90 \% cl):
\begin{eqnarray}\label{Eq:EpsBoundsDim6As}
|\varepsilon^{m,e_\mathrm{L}}_{\mu \mu}| &<& 8.2 \cdot 10^{-4}\;, \\
|\varepsilon^{m,e_\mathrm{L}}_{\tau \tau}| &<& 8.4 \cdot 10^{-3}\;, \\
\label{Eq:EpsBoundsDim6As_3}|\varepsilon^{m,e_\mathrm{L}}_{\mu \tau}| &<& 1.9 \cdot 10^{-3}\;.
\end{eqnarray}

In summary, the anti-symmetric dimension 6 operator of Eq.~(\ref{Eq:AntisymmDim6}) can only give rise to very specific NSIs (with normal matter), namely to $\varepsilon^{m,e_\mathrm{L}}_{\mu\mu}$, $\varepsilon^{m,e_\mathrm{L}}_{\mu\tau}$ and $\varepsilon^{m,e_\mathrm{L}}_{\tau\tau}$. 
The most relevant constraints come from the determination of $G_F$ via $\mu$ and $\tau$ decays (under the assumption of unitarity of the CKM matrix). Since only the shown matter NSIs involving $e_\mathrm{L}$ are generated, the bounds on $\tilde{\varepsilon}^{m}_{\alpha\beta}$ (which are defined in Eq.~(\ref{Eq:DefEpsTilde}) and which are the quantities relevant for neutrino  oscillations in matter) are the same as the bounds on the corresponding $\varepsilon^{m,e_\mathrm{L}}_{\alpha\beta}$ parameters given in Eqs.~(\ref{Eq:EpsBoundsDim6As}) - (\ref{Eq:EpsBoundsDim6As_3}). 
The bounds on $\tilde{\varepsilon}^{m}_{\alpha\beta}$ are summarised in table \ref{tab:boundsNSIMatterFromD=6}. 
 
\begin{table}
\begin{center}
\begin{tabular}{|c|c|}
\hline NSIs from $c^{d=6,as}$				& upper bound \\ \hline 
\hline $|\tilde{\varepsilon}^{m}_{\mu\mu}|$			& $8.2 \times 10^{-4}$  \\ 
\hline $|\tilde{\varepsilon}^{m}_{\mu\tau}|$ 			& $1.9 \times 10^{-3}$  \\ 
\hline $|\tilde{\varepsilon}^{m}_{\tau\tau}|$ 			& $8.4 \times 10^{-3}$   \\ \hline
\end{tabular}
\end{center}
\caption{Bounds on the NSI parameters $\tilde{\varepsilon}^{m}_{\alpha\beta}$ relevant for neutrino oscillations  which are generated from the anti-symmetric dimension 6 operator given in Eq.~(\ref{Eq:AntisymmDim6}). 
\label{tab:boundsNSIMatterFromD=6}}
\end{table}

\subsubsection{Additionally generated NSIs at the source and their constraints}

In addition to NSIs with matter, the operator of Eq.~(\ref{antisd6comp}) can also induce non-standard neutrino production at a Neutrino Factory source. The coefficients $\lambda^i_{\alpha \beta}$ can mediate the decay $
\mu \raw e \nu \bar \nu$, coupling an incoming $\mu$ with an outgoing $\bar\nu_\alpha$ with $\alpha \neq \mu$ and the outgoing $e$ with an outgoing $\nu_\beta$ with $\beta \neq e$. Thus, both the neutrino and the anti-neutrino may have non standard flavours. Defining
\begin{eqnarray}\label{Eq:NSIs_source}
{\cal L}^s_{\mathrm{NSI}} &=& 2 \sqrt{2} G_F 
\sum_{f} \varepsilon^{s}_{e\alpha,\mu\beta} 
\left[ \bar \nu_{\mathrm{L}\alpha} \gamma^\delta \nu_{\mathrm{L}\beta} \right] 
\left[ \bar l_{e} \gamma_\delta l_{\mu}\right],
\end{eqnarray}
we take into account the possibility of both neutrinos having non-standard flavours. We notice that for $\beta = \mu$, the NSI parameters $\varepsilon^{s}_{e\alpha,\mu\beta}$ reduce to the $\varepsilon^{s}_{e\alpha}$ usually considered in the literature. Eq.~(\ref{antisd6comp}) gives:
\begin{eqnarray}
\varepsilon^{s}_{e \alpha, \mu \beta}  =  
\sum_i \frac{\lambda^i_{\mu \beta} \lambda^{i*}_{e \alpha}}{\sqrt{2} G_F m_{S_i}^2} \;.
\label{epsilond6s}
\end{eqnarray}
From the bounds of Eqs.~(\ref{raredecaysd6x}) - (\ref{raredecaysd6}) we can derive the following bounds on the $\varepsilon^{s}_{e\alpha,\mu\beta}$:
\begin{eqnarray} 
|\varepsilon^{s}_{e \mu, \mu \tau}| &<& 7.5 \cdot 10^{-2}\;, \\
|\varepsilon^{s}_{e \tau, \mu e}| &<& 3.0 \cdot 10^{-2}\;, \\
|\varepsilon^{s}_{e \tau, \mu \tau}| &<& 3.5 \cdot 10^{-4}\;.
\end{eqnarray} 

As in the case of the NSIs with matter, additional bounds can be derived from the determination of $G_F$ through $\mu$ and $\tau$ decays which allows to derive bounds on the individual $\sum_i |\lambda^i_{\alpha \beta}/m_{S_i}|^2$. $\sum_i |\lambda^i_{e \mu}/m_{S_i}|^2$ is constrained through Eq.~(\ref{constr1}) and Eq.~(\ref{constr2}) can constrain $\sum_i |\lambda^i_{e \tau}/m_{S_i}|^2$. Similarly $\sum_i |\lambda^i_{\mu \tau}/m_{S_i}|^2$ can be constrained by \cite{PDG,universality}
\begin{equation}
\sqrt{\frac{G_{\tau \rightarrow \mu \nu \bar\nu}}{G_{\mu \rightarrow e \nu 
\bar\nu}}} = \frac{\sqrt{1 + \sum_i \frac{|\lambda^i_{\mu \tau}|^2}{\sqrt{2}m_{S_i}^2 G_F}}}{\sqrt{1 + \sum_i \frac{|\lambda^i_{e \mu}|^2}{\sqrt{2}m_{S_i}^2 G_F}}}=1.0002 \pm 0.0022\;,
\label{constr3}
\end{equation}
which results in the bounds:
\begin{eqnarray} 
|\varepsilon^{s}_{e \mu, \mu e}| &<& 8.2 \cdot 10^{-4}\;, \\
|\varepsilon^{s}_{e \mu, \mu \tau}| &<& 1.8 \cdot 10^{-3}\;, \\
|\varepsilon^{s}_{e \tau, \mu e}| &<& 1.9 \cdot 10^{-3}\;, \\
|\varepsilon^{s}_{e \tau, \mu \tau}| &<& 5.7 \cdot 10^{-3}\;.
\end{eqnarray} 

\subsection{Constraints on the dimension 6 operator contributing to neutrino kinetic terms}\label{Sec:D6Kin} 
\begin{figure}
   \centering
   \includegraphics[scale=0.7]{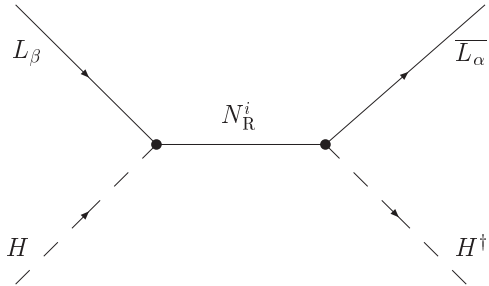}
   \caption{
   Generation of the dimension 6 operator contributing to neutrino kinetic terms by the exchange of virtual fermionic singlets $N_\mathrm{R}^i$.
   \label{fig:d=6kin}}   
\end{figure}

The second possibility to generate matter NSIs satisfying the criteria of section \ref{sec:restrictions} 
is via the dimension 6 operator
\begin{eqnarray}\label{Eq:Dim6Kin}
{\cal L}^{d=6}_{kin} =
- c^{d=6,kin}_{\alpha\beta} (\bar L_\alpha \cdot H^\dagger) \,i\cancel{\partial}\, (H \cdot L_\beta) 
\end{eqnarray} 
which induces non-canonical neutrino kinetic terms. 
After diagonalising and normalising the neutrino kinetic terms, 
a non-unitary lepton mixing matrix is produced from this operator.
The tree level generation of this operator, avoiding a similar contribution to charged leptons that would lead to flavour changing neutral currents, requires the introduction of SM-singlet fermions (right-handed neutrinos) which couple to the Higgs and lepton doublets via the Yukawa couplings (see e.g.\ \cite{Abada:2007ux}) as shown in figure \ref{fig:d=6kin}, 
\begin{equation}\label{Eq:EffOpInTheMiddle}
 \mathcal{L}_{int}^Y = -Y^*_{\alpha i}  (\bar{L}_{\alpha} \cdot  H^{\dagger}) N_\mathrm{R}^{i} + \text{H.c.}\;.
\end{equation}

When singlet fermions (right-handed neutrinos) with Yukawa couplings and a (Majorana) mass matrix are introduced, this can in general lead to two effective operators at tree-level: It can, on the one hand, produce the dimension 5 neutrino mass operator (Weinberg operator) \cite{Weinberg:1979sa} which generates neutrino masses after EW symmetry breaking and violates lepton number. On the other hand, this extension of the SM always leads to the dimension 6 operator of Eq.~(\ref{Eq:Dim6Kin}) which contributes to the kinetic energy of the neutrinos and induces non-unitarity of the leptonic mixing matrix. As we will discuss in section \ref{sec:ConstraintsTypeB}, the constraints on the diagonal elements of this dimension 6 operator can be used to constrain the NSIs induced by the dimension 8 operator. 
We note that the dimension 5 (Weinberg) operator for neutrino masses does not lead to additional constraints, because it can be suppressed by an approximate global U(1) ``lepton number'' symmetry \cite{Bernabeu:1987gr,GonzalezGarcia:1988rw,Tommasini:1995ii,Pilaftsis:2004xx,Pilaftsis:2005rv,Kersten:2007vk,Abada:2007ux}. 
The smallness of neutrino masses in this case is not explained by large masses of singlet fermions but by the smallness of lepton number breaking effects. Sizable non-unitarity of the leptonic mixing matrix could arise from comparatively light singlet fermions (with un-suppressed Yukawa couplings), without being in conflict with the smallness of the neutrino masses.  
Generally speaking, both operators are not directly related since the dimension 5 neutrino mass operator violates lepton number while the dimension 6 operator is lepton number conserving. 

From the point of view of neutrino oscillation experiments, having in mind in particular a possible future Neutrino Factory, we will regard right-handed neutrinos with masses  $M_{i}$ above a few GeV as ``heavy'', such that we can effectively integrate them out of the theory.\footnote{Notice that even if $M_i < v$ it will turn out that $\left| v Y_{\alpha i}/M_i \right| \ll 1$, such that an effective operator expansion is still possible.}   
In the following, we review and update the constraints derived in \cite{Antusch:2006vwa} on the product $NN^\dagger$ (where $N$ is the non-unitary lepton mixing matrix) for $M_{i}$ larger than the EW scale $\Lambda_{\mathrm{EW}}$, and extend the constraints to $M_{i}$ larger than a few GeV but below $\Lambda_{\mathrm{EW}}$. In the following, without loss of generality, we will always work in the basis where the charged lepton Yukawa matrix is diagonal.

\subsubsection{The case $\boldsymbol{M_{i}$ above $\Lambda_{\mathrm{EW}}}$}
In \cite{Antusch:2006vwa} the diagonal elements of $NN^\dagger$ were constrained by the combination of universality tests and the invisible decay width of the $Z$. Notice that, without the inclusion of the invisible width of the $Z$, all the constraints derived would consist of ratios of elements of $NN^\dagger$ and an uncertainty on their overall scale would remain. This can be easily understood from the fact that in the Lagrangian the mixing matrix $N$ is always multiplied by the weak coupling constant $g$ and, since $G_F$ is measured through the $\mu$ decay, the comparison of any leptonic process will lead to ratios of the elements of $NN^\dagger$. Apart from the invisible width of the $Z$, this can be resolved by comparing leptonic and hadronic processes as in section \ref{CKM}. Indeed the extraction of the Fermi constant from the $\mu$ decay with non-unitary leptonic mixing leads to
\begin{equation}
\label{Eq:GFM}
G_\mu = G_F \sqrt{(NN^\dagger)_{ee} (NN^\dagger)_{\mu\mu} }\;.
\end{equation}
Performing the steps as in section \ref{CKM}, we obtain
\begin{equation}
|V_{ud}^{exp}|^2 + |V_{us}^{exp}|^2  = 
 \frac{1}{(NN^\dagger)_{\mu\mu} }  = 0.9997 \pm 0.0010 \;.
\end{equation}
To update the bounds of \cite{Antusch:2006vwa}, we replace the bound from the invisible decay width of the $Z$ by this more tight constraint.

Furthermore, the off-diagonal elements of $NN^\dagger$ are constrained by rare radiative lepton decays, $l_\alpha \raw l_\beta \gamma$. With respect to the bounds derived in \cite{Antusch:2006vwa} we also add here the contribution of the diagrams mediated by the heavy right-handed neutrinos. This was not considered in \cite{Antusch:2006vwa} where a more model independent approach to the source of non-unitarity (based on the so-called Minimal Unitarity Violation scheme (MUV) where an extension of the SM by only the dimension 5 Weinberg operator and the dimension 6 operator of Eq.~(\ref{Eq:Dim6Kin}) is considered) was adopted. 
Notice also that the constraints on the diagonal elements can be used to obtain bounds on the off-diagonal ones when the former are stronger, using:

\begin{eqnarray}
\frac{v^2}{2}|c^{d=6,kin}_{\alpha\beta}| &=& \frac{v^2}{2} \left| \sum_i \frac{Y^*_{\alpha i} Y_{\beta i}}{M^2_i} \right| \leq \frac{v^2}{2} \sqrt{\sum_i \left| \frac{Y_{\alpha i}}{M_i} \right|^2 \sum_j \left|\frac{Y_{\beta j}}{M_j} \right|^2} \nonumber \\
&=& \frac{v^2}{2} \sqrt{|c^{d=6,kin}_{\alpha\alpha}||c^{d=6,kin}_{\beta\beta}|}\;.
\label{Schwartzid}
\end{eqnarray}
In combination with the additional constraints considered in \cite{Antusch:2006vwa}, we obtain the following updated bounds at 90\% cl:
\begin{eqnarray}\label{Eq:D=6MaboveEW}
|(N N^\dagger)_{\alpha \beta} - \delta_{\alpha \beta}| = \frac{v^2}{2}|c^{d=6,kin}_{\alpha\beta}| < 
\begin{pmatrix}
4.0 \cdot 10^{-3}  & 1.2 \cdot 10^{-4}  &  3.2 \cdot 10^{-3}\\
1.2 \cdot 10^{-4}  & 1.6 \cdot 10^{-3}  &  2.1 \cdot 10^{-3}\\
3.2 \cdot 10^{-3}  & 2.1 \cdot 10^{-3}  &  5.3 \cdot 10^{-3} 
\end{pmatrix} \, .
\label{nndag}
\end{eqnarray}

\subsubsection{The case $\boldsymbol{M_{i}$ below $\Lambda_{\mathrm{EW}}$ but above a few GeV$}$}
One might think that the constraints on the generated NSIs could be significantly relaxed if the singlet fermions are lighter than $\Lambda_{\mathrm{EW}}$. For completeness, we will therefore also discuss the situation where 
$M_{i}$ is below $\Lambda_{\mathrm{EW}}$ but above a few GeV. From the point of view of neutrino oscillation experiments, right-handed neutrinos below $\Lambda_{\mathrm{EW}}$ but above the typical energies of the experiment can still be considered as heavy (and can thus be effectively integrated out of the theory inducing non-unitarity of the leptonic mixing matrix). 
In general, the constraints on non-unitarity of the leptonic mixing matrix from the decays of particles with masses above the masses $M_{i}$ of the  right-handed neutrinos are indeed lost, since all the mass eigenstates are now available in the decay and unitarity is restored. Thus, the $Z$ and $W$ decays cannot be used anymore, however the constraints on the diagonal elements of $N N^\dagger$ derived from $\mu$ decays, $\beta$ decays and kaon decays together with the universality constraints from $\tau$ and $\pi$ decays still apply and these can still be translated into bounds on the off-diagonal elements using Eq.~(\ref{Schwartzid}). Only the strong constraint on the $e \mu$ element from $\mu \raw e \gamma$ is lost due to the restoration of the GIM mechanism. In summary we obtain the following bounds (at 90\% cl): 
\begin{eqnarray}\label{Eq:D=6MbelowEW}
|(N N^\dagger)_{\alpha \beta} - \delta_{\alpha \beta}| = \frac{v^2}{2}|c^{d=6,kin}_{\alpha\beta}| < 
\begin{pmatrix}
4.0 \cdot 10^{-3}  & 1.8 \cdot 10^{-3}  &  3.2 \cdot 10^{-3}\\
1.8 \cdot 10^{-3}  & 1.6 \cdot 10^{-3}  &  2.1 \cdot 10^{-3}\\
3.2 \cdot 10^{-3}  & 2.1 \cdot 10^{-3}  &  5.3 \cdot 10^{-3} 
\end{pmatrix} \, .
\label{nndag_low}
\end{eqnarray}

\subsubsection{NSIs with matter induced by the dimension 6 operator which contributes to neutrino kinetic terms}\label{Sec:ConstraintsD=6Kin}
As discussed above, the dimension 6 operator which contributes to neutrino kinetic terms leads to non-unitarity of the leptonic mixing matrix, i.e.\ to $(N N^\dagger)_{\alpha\beta} \not= \delta_{\alpha\beta}$. Therefore (c.f.\ \cite{Antusch:2006vwa}), it gives rise to non-standard matter interactions as well as to non-standard interactions at the source and detector, which are related to the matter NSIs. In the following, we will review the bounds on the matter NSIs in this case. However, we would like to emphasise that the related non-standard interactions at the source and detector may also have  strong (or even stronger) effects on neutrino oscillation experiments.\footnote{The formalism for a full treatment of neutrino oscillations in the presence of such non-unitarity of the leptonic mixing matrix can be found in \cite{Antusch:2006vwa}. The NSI parameterisation of new physics in neutrino oscillations can also be applied to the case of non-unitarity. Using the NSI approach takes account of the leading order effects of the modified interaction with the $W$ and $Z$ bosons, which are induced by the dimension 6 operator in Eq.~(\ref{Eq:Dim6Kin}) after EW symmetry breaking and canonically normalising the neutrino kinetic terms.}

Using the bounds in Eqs.~(\ref{Eq:D=6MaboveEW}) and (\ref{Eq:D=6MbelowEW}) and taking into account that the  interactions with the $W$ and $Z$ bosons are modified to $N_{\alpha i}$ and $(N N^\dagger)_{\alpha\beta}$, respectively, we can compute the bounds on the individual NSI parameters in matter induced by the dimension 6 operator of Eq.~(\ref{Eq:Dim6Kin}) using the relations:
\begin{eqnarray}
\varepsilon^{m,e_\mathrm{L}}_{\alpha\beta} &=&
-\frac{1}{2} \left(\frac{v^2}{2} c^{d=6,kin}_{\alpha e} \delta_{\beta e} + \frac{v^2}{2} c^{d=6,kin}_{e \beta}  \delta_{e \alpha}\right) + \left(\frac{1}{2} - \sin^2 \theta_W\right) \,\frac{v^2}{2}c^{d=6,kin}_{\alpha \beta}\;,\\
\varepsilon^{m,e_\mathrm{R}}_{\alpha\beta} &=& - \sin^2 \theta_W \,\frac{v^2}{2}c^{d=6,kin}_{\alpha \beta}\;,\\
\varepsilon^{m,u_\mathrm{L}}_{\alpha\beta} &=&
- \left(\frac{1}{2} - \frac{2}{3}\sin^2 \theta_W\right) \,\frac{v^2}{2}c^{d=6,kin}_{\alpha \beta}\;,\\
\varepsilon^{m,u_\mathrm{R}}_{\alpha\beta} &=&
\frac{2}{3}\sin^2 \theta_W \,\frac{v^2}{2}c^{d=6,kin}_{\alpha \beta}\;,\\
\varepsilon^{m,d_\mathrm{L}}_{\alpha\beta} &=&
 \left(\frac{1}{2} - \frac{1}{3}\sin^2 \theta_W \right) \,\frac{v^2}{2}c^{d=6,kin}_{\alpha \beta}\;,\\
\varepsilon^{m,d_\mathrm{R}}_{\alpha\beta} &=&
 - \frac{1}{3}\sin^2 \theta_W \,\frac{v^2}{2}c^{d=6,kin}_{\alpha \beta}\;.
\end{eqnarray}
Using these relations the parameters $\tilde{\varepsilon}^m_{\alpha\beta}$ defined in Eq.~(\ref{Eq:DefEpsTilde}) are given by (see e.g.\ \cite{Campanelli:2002cc})
\begin{eqnarray}
\tilde{\varepsilon}^m_{\alpha\beta} = -\frac{1}{2} \left(\frac{v^2}{2} c^{d=6,kin}_{\alpha e} \delta_{\beta e} + \frac{v^2}{2} c^{d=6,kin}_{e \beta}  \delta_{e \alpha}\right) + \frac{1}{2} \frac{n_n}{n_e} \,\left(\frac{v^2}{2} c^{d=6,kin}_{\alpha \beta}\right) ,
\end{eqnarray}
which leads to the constraints
\begin{eqnarray}
|\tilde{\varepsilon}^m_{\alpha\beta}| < \frac{v^2}{2}\!
\begin{pmatrix}
|\frac{1}{2}\left(\frac{n_n}{n_e}-2\right)\,c^{d=6,kin}_{e e}| & |\frac{1}{2}\left(\frac{n_n}{n_e}-1\right)\,c^{d=6,kin}_{e \mu}|& |\frac{1}{2}\left(\frac{n_n}{n_e}-1\right)\,c^{d=6,kin}_{e \tau}|\\
|\frac{1}{2}\left(\frac{n_n}{n_e}-1\right)\,c^{d=6,kin}_{e \mu}| & |\frac{1}{2} \frac{n_n}{n_e}\,c^{d=6,kin}_{\mu \mu}|& |\frac{1}{2} \frac{n_n}{n_e}\,c^{d=6,kin}_{\mu \tau}|\\
|\frac{1}{2}\left(\frac{n_n}{n_e}-1\right)\,c^{d=6,kin}_{e \tau}| & |\frac{1}{2} \frac{n_n}{n_e}\,c^{d=6,kin}_{\mu \tau}|&|\frac{1}{2} \frac{n_n}{n_e}\,c^{d=6,kin}_{\tau \tau}|
\end{pmatrix}
\end{eqnarray}
with $\frac{v^2}{2} c^{d=6,kin}_{\alpha \beta}$ replaced by their upper bounds given in Eqs.~(\ref{Eq:D=6MaboveEW}) and (\ref{Eq:D=6MbelowEW}) for $M_i$ above or below $\Lambda_\mathrm{EW}$, respectively. Since the ratio $\frac{n_n}{n_e}$ is in general close to $1$, this implies that the bounds on $|\tilde{\varepsilon}^m_{e \mu}|$ and $|\tilde{\varepsilon}^m_{e \tau}|$ are significantly stronger than the bounds on the individual $\varepsilon^{m,f}_{\alpha\beta}$. In table \ref{Tab:nnoverne} the values of the \% weight amount of the main constituents of the Earth's continental crust \cite{crust} and mantle \cite{mantle} together with the mean value of $\frac{n_n}{n_e}$ inferred from that composition are given. Notice that the factor $\frac{n_n}{n_e}-1$ means an additional suppression of two orders of magnitude of the NSI coefficient.

\begin{table}
\begin{center}
\begin{tabular}{|c|c|}
\hline NSIs from $c^{d=6,kin}$	& upper bound (for $M_i > $ few GeV) \\ \hline
\hline $|\tilde{\varepsilon}^{m}_{ee}|$		& $2.0 \times 10^{-3} \times |\frac{n_n}{n_e}-2|$  \\ 
\hline $|\tilde{\varepsilon}^{m}_{e\mu}|$ 	& $9.1 \times 10^{-4} \times |\frac{n_n}{n_e}-1|$  \\
 & (for $M_i \gg \Lambda_\mathrm{EW}$: $5.9 \times 10^{-5} \times |\frac{n_n}{n_e}-1|$)\\ 
\hline $|\tilde{\varepsilon}^{m}_{e\tau}|$ 	& $1.6 \times 10^{-3} \times |\frac{n_n}{n_e}-1|$  \\ 
\hline $|\tilde{\varepsilon}^{m}_{\mu\mu}|$	& $8.2 \times 10^{-4} \times \frac{n_n}{n_e}$  \\ 
\hline $|\tilde{\varepsilon}^{m}_{\mu\tau}|$ 	& $1.0 \times 10^{-3} \times \frac{n_n}{n_e}$  \\ 
\hline $|\tilde{\varepsilon}^{m}_{\tau\tau}|$ 	& $2.6 \times 10^{-3} \times \frac{n_n}{n_e}$  \\ 
\hline
\end{tabular}
\end{center}
\caption{ Bounds on the NSI parameters $\tilde{\varepsilon}^{m}_{\alpha\beta}$ relevant for neutrino oscillations which are generated from the dimension 6 operator which contributes to the neutrino kinetic terms, given in Eq.~(\ref{Eq:Dim6Kin}). Values $n_n / n_e$ for the crust and the mantle of the earth can be found in table \ref{Tab:nnoverne}. 
\label{tab:boundsNSIMatterFromD=6Kin}}
\end{table}

\begin{table}
\begin{center}
\begin{tabular}{|c|c|c|}
\hline
Compound  & Crust & Mantle \\ \hline
\hline SiO$_2$	   & 60.6 & 46.0 \\
\hline Al$_2$O$_3$ & 15.9 & 4.2 \\
\hline FeO & 6.7 & 7.5 \\
\hline CaO & 6.4 & 3.2 \\
\hline MgO & 4.7 & 37.8 \\
\hline Na$_2$O & 3.1 & 0.4 \\
\hline K$_2$O & 1.8 & 0.04 \\ \hline
\hline $n_n / n_e$ & 1.017 & 1.019 \\ \hline
\end{tabular}
\end{center}
\caption{Values of the $\%$ weight amount of the main constituents of the earth's continental crust \cite{crust} and mantle \cite{mantle} together with the mean value of $\frac{n_n}{n_e}$ inferred from that composition. 
\label{Tab:nnoverne}}
\end{table}

\subsubsection{Additionally generated NSIs at the source and at the detector and constraints}
A non-unitary neutrino mixing matrix $N$ leads to non-standard interactions at the source and at the detector of a neutrino oscillation experiment due to the modified coupling to the $W$. In Ref.~\cite{FernandezMartinez:2007ms} it has been shown that, parameterising the non-unitary matrix as $N = (1 + \eta)\,U$ where $\eta$ is a (small) Hermitian matrix and U is unitary,
the NSI coefficients at the source and detector can be expressed in terms of $\eta$ as $\varepsilon^{s}_{\alpha \beta} = \varepsilon^{d}_{\alpha \beta} = \eta_{\alpha \beta}$. Since $NN^\dagger = (1 + \eta)^2 \simeq 1+2\eta$, the bounds on Eqs.~(\ref{nndag}) and (\ref{nndag_low}) can be translated into  bounds on 
\begin{eqnarray}
\varepsilon^{s}_{\alpha \beta} = \varepsilon^{d}_{\alpha \beta} = \tfrac{1}{2}((N N^\dagger)_{\alpha \beta} - \delta_{\alpha \beta}) = \frac{v^2}{4} c^{d=6,kin}_{\alpha\beta}\;.
\end{eqnarray}

\section{Dimension 8 operators for matter NSIs}\label{Sec:Dim8}
We now consider the possibility of generating NSIs with matter from dimension 8 operators under the restrictions discussed in section \ref{sec:restrictions}. In particular, we analyse operators with 
$L,\bar L, f, \bar f, H, H^\dagger$ as external fields, where $f$ can be $f_\mathrm{L}$ or $f^c_\mathrm{R}$   with $f_\mathrm{L} \in \{L_1,Q_1\}$ and $f^c_\mathrm{R} \in \{e^c_\mathrm{R},u^c_\mathrm{R},d^c_\mathrm{R}\}$
including their generation at tree-level.

\begin{figure}
 \centering  
 \subfigure[Topology 1]{\includegraphics[scale=1]{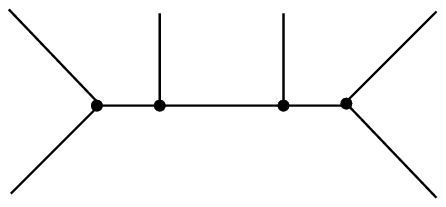}}
 \quad\quad
 \subfigure[Topology 2]{\includegraphics[scale=1]{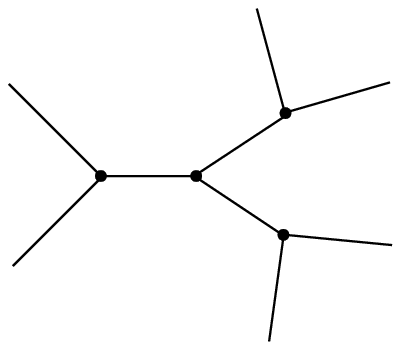}}
 \quad\quad
 \subfigure[Topology 3]{\includegraphics[scale=1]{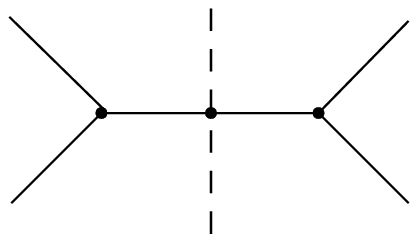}}
 \caption{\label{fig:topologies}
Topologies of tree-level Feynman diagrams which can realise the relevant dimension 8 operators. The solid external lines in diagrams (a) and (b) correspond to the fields $L,\bar L, f, \bar f, H, H^\dagger$ (where $f$ can be $f_\mathrm{L}$ or $f^c_\mathrm{R}$  with $f_\mathrm{L} \in \{L_1,Q_1\}$ and $f^c_\mathrm{R} \in \{e^c_\mathrm{R},u^c_\mathrm{R},d^c_\mathrm{R}\}$). In diagram (c) the dashed lines indicate the SM Higgs fields $H$ and $H^\dagger$. 
 }
\end{figure}

To start with, operators with these external fields can be generated at tree-level via the three topologies of Feynman diagrams shown in figure \ref{fig:topologies}. Scanning over these topologies and possibilities for the external fields to be arranged (up to here there are 51 inequivalent diagrams for topology 1, 11 for topology 2 and 3 for topology 3) as well as over the virtual states allowed to be interchanged, we find only three classes of possibilities which satisfy the criteria of section \ref{sec:restrictions}.\footnote{We note that in some of the discarded cases the interactions of four charged fermions appear after canonically normalising the kinetic terms of the fields.}  
We will now discuss these three cases as well as the corresponding constraints on the NSIs with matter.

\subsection{Case I: Coupling fermionic singlets to lepton and Higgs doublets} \label{sec:ConstraintsTypeB}
\begin{figure}
 \centering  
 \subfigure[ ]{\includegraphics[scale=0.7]{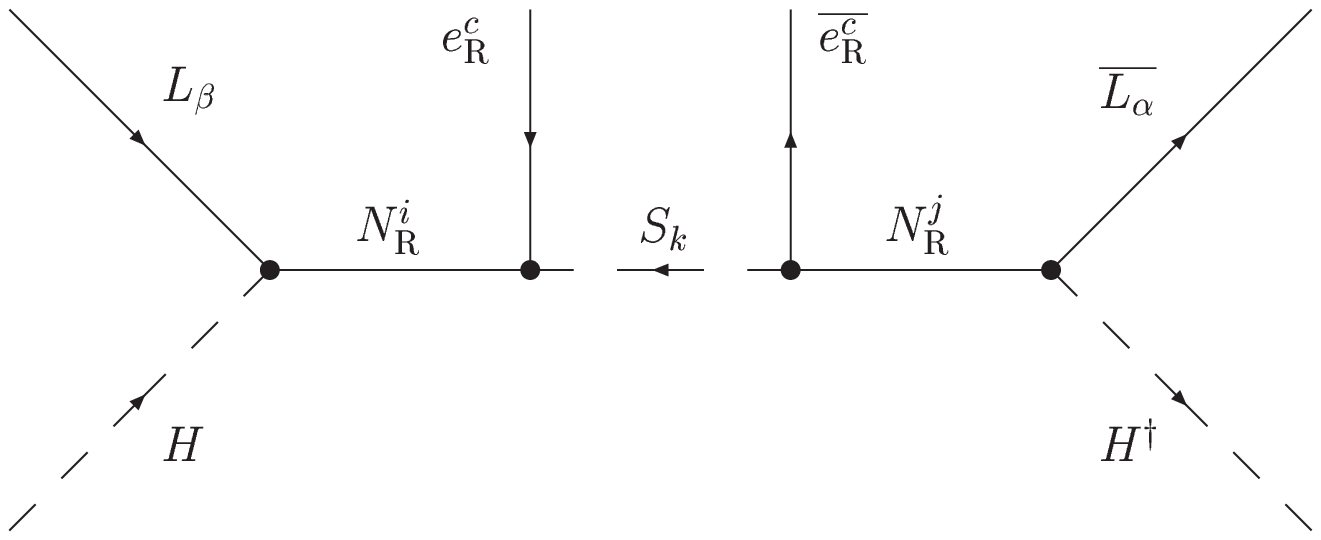}}\\[1cm]
 \subfigure[ ]{\includegraphics[scale=0.7]{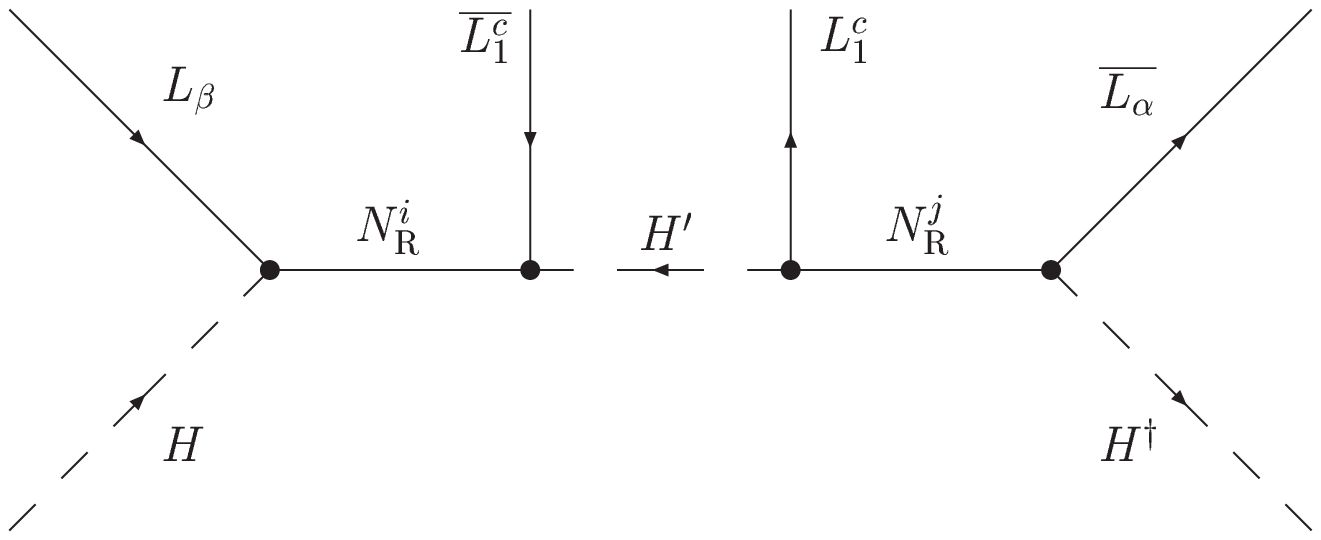}}\\[1cm]
 \subfigure[ ]{\includegraphics[scale=0.7]{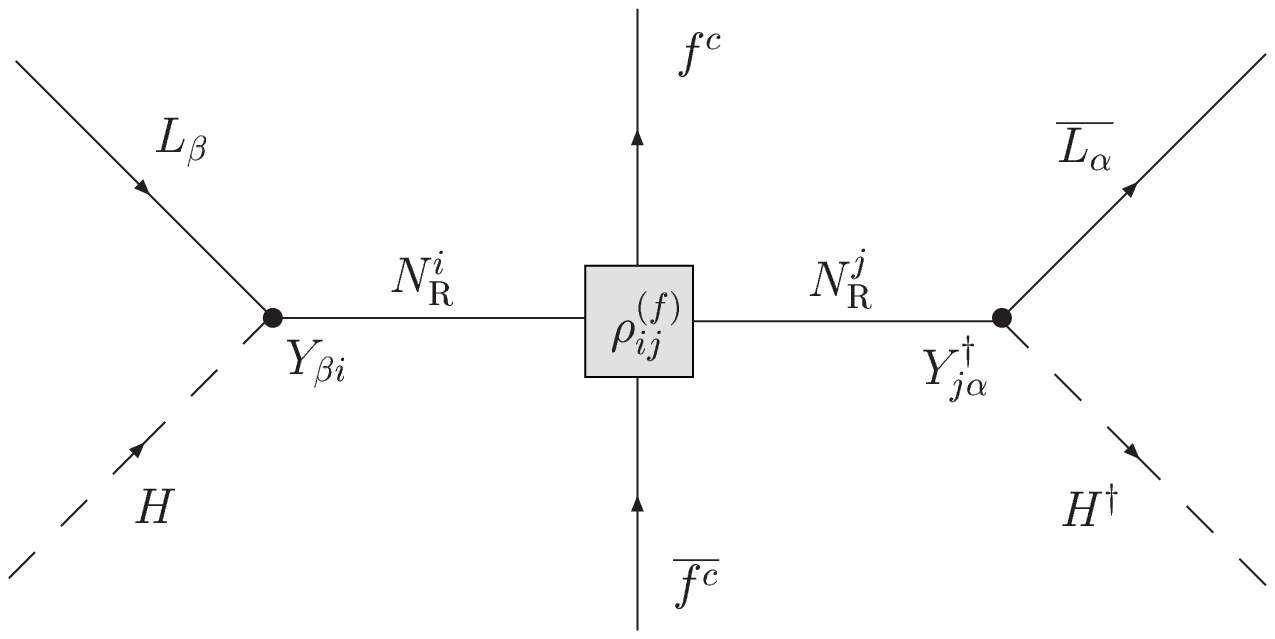}}
 \caption{\label{fig:d=8case1}
Diagrams (a) and (b): Generation of dimension 8 operators which induce NSIs with matter after EW symmetry breaking in extensions of the SM with fermionic singlets $N_\mathrm{R}^i$ coupling to lepton and Higgs doublets (via topology 1 of figure \ref{fig:topologies}). Similar diagrams exist with quarks as external fields.  
Diagram (c): Feynman diagram which summarises the diagrams (a), (b) as well as similar diagrams with quarks, introduced in order to simplify the discussion of constraints on the generated NSIs. $\rho_{ij}^{(f)}$ are effective operators defined in Eq.~(\ref{Eq:EffOpInMiddle}). $f$ stand for $f_\mathrm{L}$ or $f^c_\mathrm{R}$  with $f_\mathrm{L} \in \{L_1,Q_1\}$ and $f^c_\mathrm{R} \in \{e^c_\mathrm{R},u^c_\mathrm{R},d^c_\mathrm{R}\}$.
 }
\end{figure}

One generic possibility to generate NSIs with matter via dimension 8 operators while satisfying the criteria of section \ref{sec:restrictions} is to couple two pairs of lepton and Higgs doublets to SM singlet fermions (right-handed neutrinos) $N_\mathrm{R}^i$, via the Yukawa interactions of Eq.~(\ref{Eq:EffOpInTheMiddle}).
Examples for diagrams where the external fields $f$ are either $e^c_{\mathrm{R}}$ or $L_1$ are shown in figure \ref{fig:d=8case1}(a) and (b). Similar diagrams exist with quarks as external fields. 
The resulting dimension 8 operators have the form \cite{Davidson:2003ha,BerezhianiRossi01}
\begin{eqnarray}\label{Eq:Dim8TypeB}
{\cal L}^{d=8,I}_{NSI} =
 c^{d=8,f,I}_{\alpha\beta}  (\overline{L}_\alpha \cdot H^\dagger) \,f^c \overline{f^c}\, (H \cdot  L_\beta) 
\;.
\end{eqnarray}   
These operators are often quoted as examples how to realise very large non-standard matter effects.
To summarise the classes of diagrams which realise the dimension 8 operators of Eq.~(\ref{Eq:Dim8TypeB}), we introduce another effective non-renormalisable operator (c.f.\ figure \ref{fig:d=8case1}(c)), 
\begin{equation}\label{Eq:EffOpInMiddle}
 \mathcal{L}_{int}^{\rho,f} = \rho^{(f)}_{ij}  \overline{N_\mathrm{R}}_{i}  f \overline{f}  N_{\mathrm{R}j}   \;.
\end{equation}
In the examples in figure \ref{fig:d=8case1} this operator is generated by the exchange of a virtual singly charged scalar field or by an inert Higgs doublet (which does not get a vacuum expectation value). 
In the following, we will assume that the right-handed neutrinos are heavier than the typical scale of a neutrino experiment, such that they can be effectively integrated out of the theory and the dimension 8 operators of Eq.~(\ref{Eq:Dim8TypeB}) remain.

To derive constraints on the NSIs generated by the operators of Eq.~(\ref{Eq:Dim8TypeB}), we first match
the ``full theory'' with the dimension 8 operators. This leads to the following relation for the coefficients $c^{d=8,f,I}_{\alpha\beta}$:
\begin{eqnarray}
c^{d=8,f,I}_{\alpha\beta} = 
\sum_{ij} \frac{Y_{\beta  i}}{M_i}  
\rho^{(f)}_{ij} 
\frac{Y^{\dagger}_{j \alpha}}{M_j}\;.
\end{eqnarray}
The corresponding NSI parameters are given by 
\begin{eqnarray}
|\varepsilon^{m,f}_{\alpha\beta}| = \left|\frac{v^2\, c^{d=8,f}_{\alpha\beta}}{4 \sqrt{2} G_F}\right| = 
\frac{v^2}{4 \sqrt{2}} 
\left|\sum_{ij} \frac{Y_{\beta  i}}{M_i}\frac{\rho^{(f)}_{ij}}{G_F} \frac{Y^{\dagger}_{j \alpha}}{M_j}\right|.
\end{eqnarray}
We continue by noting that
\begin{eqnarray}\label{Eq:ConstraintsDim8_B_1}
\frac{v^2}{2} \left|\sum_{ij}  
\frac{Y_{\beta  i}}{M_i} 
\frac{\rho^{(f)}_{ij}}{G_F} 
\frac{Y^{\dagger}_{j\alpha}}{M_j}\right| 
&\leq& \frac{v^2 \hat\rho^{(f)}}{2 G_F} \sqrt{\sum_i \left| \frac{Y_{\beta i}}{M_i} \right|^2 \sum_j \left|\frac{Y_{\alpha j}}{M_j} \right|^2} \nonumber \\
& =& \frac{v^2 \hat\rho^{(f)}}{2 G_F} \sqrt{|c^{d=6,kin}_{\beta\beta}||c^{d=6,kin}_{\alpha\alpha}|} \;,
\end{eqnarray} 
where $\hat\rho^{(f)}$ is the modulus of the largest eigenvalue of $\rho^{(f)}_{ij}$. 
The dimension 8 operators of Eq.~(\ref{Eq:Dim8TypeB}) thus turn out to be constrained by the bounds on the dimension 6 operator contributing to neutrino kinetic energy.

Now we can use the constraints from the dimension 6 operator contributing to neutrino kinetic energy given in Eq.~(\ref{nndag}).   
This leads to the bounds (at 90\% cl)
\begin{eqnarray}\label{Eq:ConstraintsD=8}
|\varepsilon^{m,f}_{\alpha\beta}| < 
 \begin{pmatrix}
1.4 \cdot 10^{-3}  & 6.4 \cdot 10^{-4}  &  1.1 \cdot 10^{-3}\\
6.4 \cdot 10^{-4}  & 5.8 \cdot 10^{-4}  &  7.3 \cdot 10^{-4}\\
1.1 \cdot 10^{-3}  & 7.3 \cdot 10^{-4}  &  1.9 \cdot 10^{-3} 
\end{pmatrix} \frac{\hat\rho^{(f)}}{G_F}\, ,
\end{eqnarray} 
which applies to both left- and right-handed fermions. 

The bounds of Eq.~(\ref{Eq:ConstraintsD=8}) depend on the quantity $\hat\rho^{(f)}/G_F$ which gives the relative strength of the effective coupling of Eq.~(\ref{Eq:EffOpInMiddle}) with respect to the Fermi constant. Let us now quantify how large this factor could be: The extra particles beyond the SM required to generate this effective coupling are scalar fields which contain (after EW symmetry breaking) an electrically charged component. Such electrically charged scalars would have been produced by pairs via photon or $Z$ exchange in the s channel by LEP if their masses were lower than $\sim 70$ GeV \cite{PDG}. For NSIs with quarks, $e_\mathrm{R}$ and $L_1$ should be replaced by their quark counterparts in figure 
\ref{fig:d=8case1}(a) and (b). The extra scalar fields required in this case would additionally be colored and much more tightly constrained.
Taking the $70$ GeV bound and, for example, couplings of the scalar fields to the right-handed neutrinos of 
${\cal O}(1)$, we find that  $\rho^{(f)}_{ij}/G_F \lesssim 10$. The bounds on the NSI parameters $|\varepsilon^{m,f}_{\alpha\beta}|$ are thus at least ${\cal O}(10^{-2})$.

Furthermore, we would like to emphasise that although the dimension 8 operator of Eq.~(\ref{Eq:Dim8TypeB}) itself produces only NSIs with matter, the additionally generated dimension 6 operator also gives rise to non-standard interactions with matter (c.f.\ discussion in section \ref{Sec:ConstraintsD=6Kin}), and additionally it gives rise to non-standard interactions at the source and detector.

\subsection{Case II: Coupling singly charged scalars to lepton doublets} \label{sec:ConstraintsTypeB2}
\begin{figure}
 \centering  
 \subfigure[ ]{\includegraphics[scale=0.7]{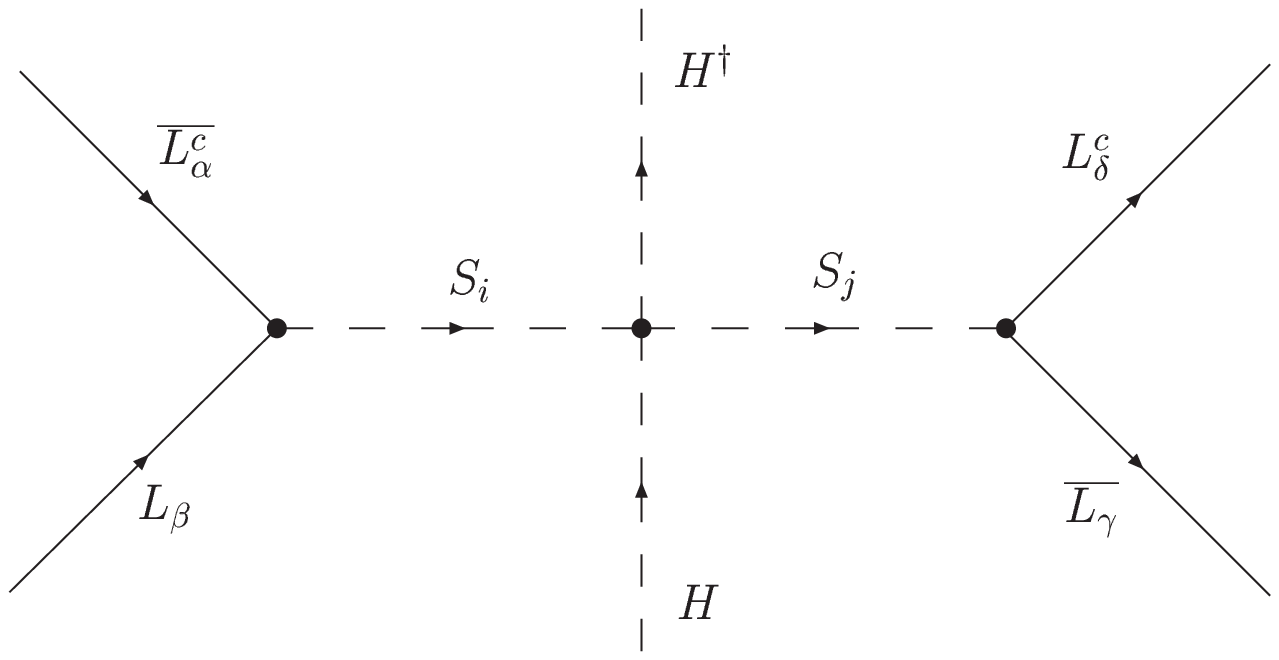}}\\[1cm]
 \subfigure[ ]{\includegraphics[scale=0.7]{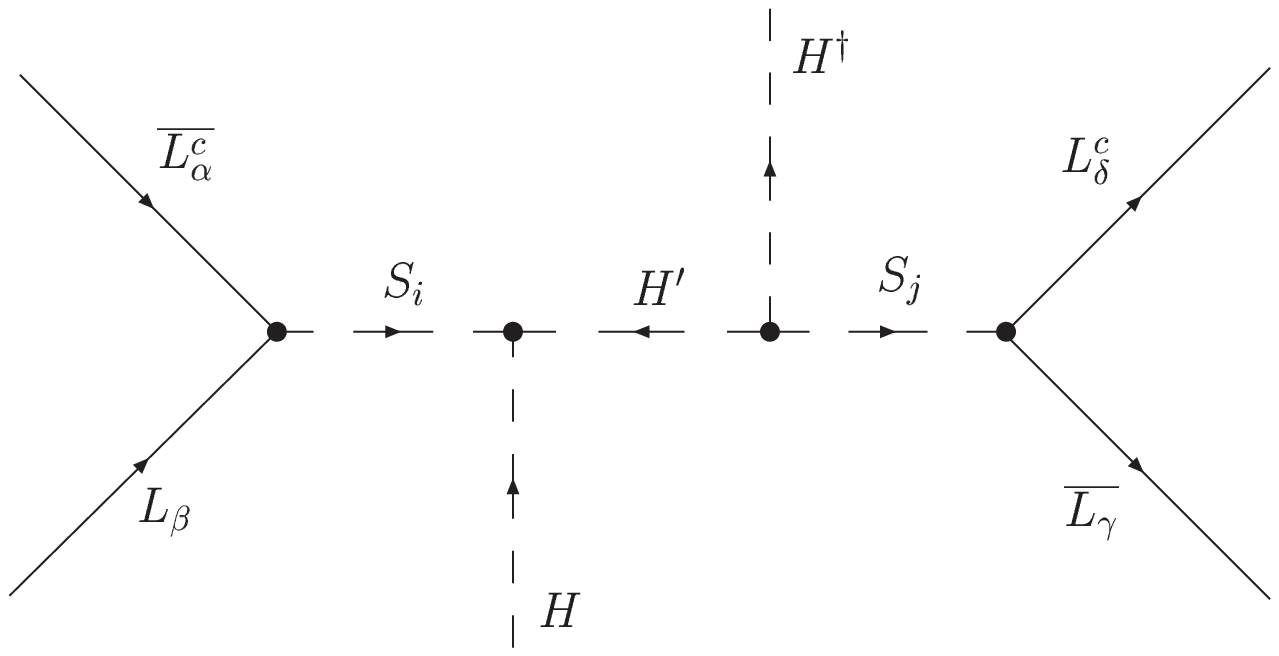}}\\[1cm]
 \subfigure[ ]{\includegraphics[scale=0.7]{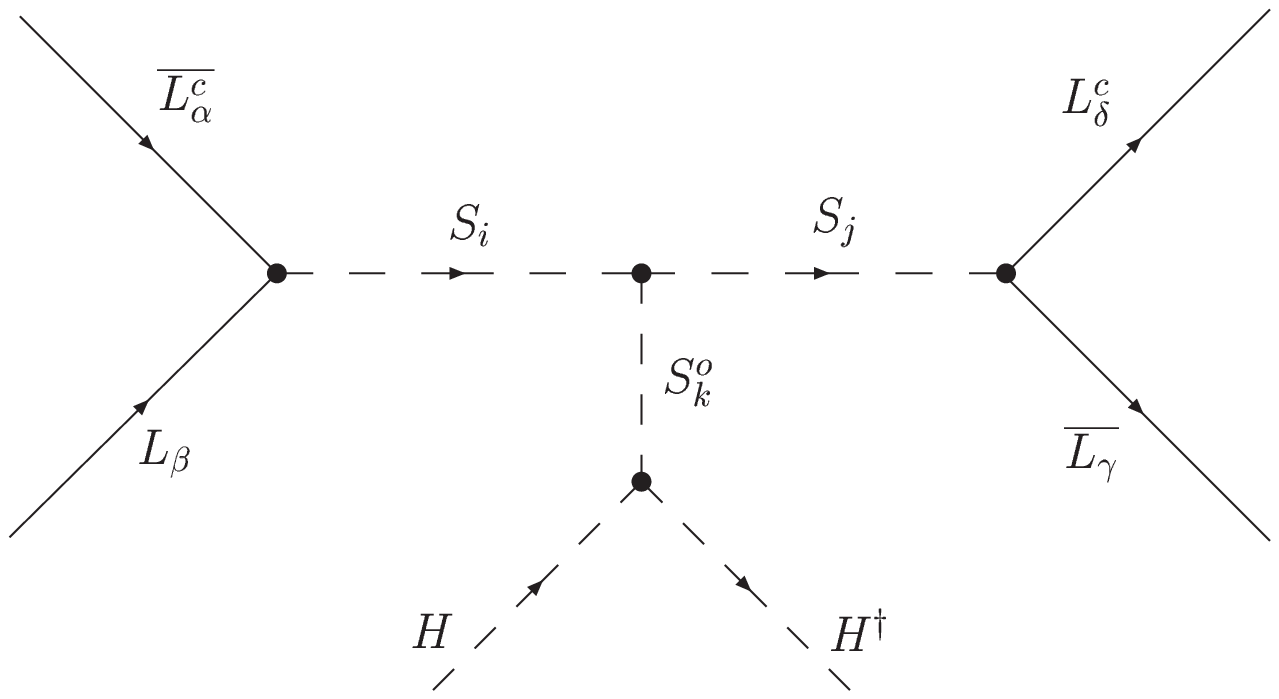}}
 \caption{\label{fig:d=8case2}
Generation of dimension 8 operators which induce NSIs with matter after EW symmetry breaking in extensions of the SM with antisymmetric couplings of singly charged scalars $S_i$ to two lepton doublets (via topologies 1, 2 and 3 of figure \ref{fig:topologies}). Effectively, the two couplings to the Higgs field only contribute to the propagators of the $S_i$ after EW symmetry breaking and consequently the constraints on the NSIs in this case are the same as for the anti-symmetric dimension 6 operator in section \ref{App:AnitSymmDim6}. 
 }
\end{figure}

Another generic possibility to select one neutrino and one charged lepton (and to avoid generating in addition couplings between two charged leptons) is to couple a pair of lepton doublets to singly charged scalar fields $S_i$, as in Eq.~(\ref{Eq:IntS}). The possible diagrams, which  
generate dimension 8 operators of the form
\begin{eqnarray}\label{Eq:Dim8CaseII}
{\cal L}^{d=8}_{NSI,II} =
c^{d=8,f,II}_{\alpha\beta\gamma\delta} (\overline{L^c}_\alpha \cdot L_\beta) (\bar L_\gamma \cdot L^c_\delta) (H^\dagger H)
\;,
\end{eqnarray}
are shown in figure \ref{fig:d=8case2}. 
For NSIs from these dimension 8 operators, the same constraints as the ones for NSIs from the anti-symmetric dimension 6 operator (c.f.\ section \ref{App:AnitSymmDim6}) still apply, since after EW symmetry breaking the dimension 6 operator is recovered.

\subsection{Case III: Mixed case with singly charged scalars and singlet fermions} \label{sec:ConstraintsTypeB3}
\begin{figure}
 \centering  
\includegraphics[scale=0.7]{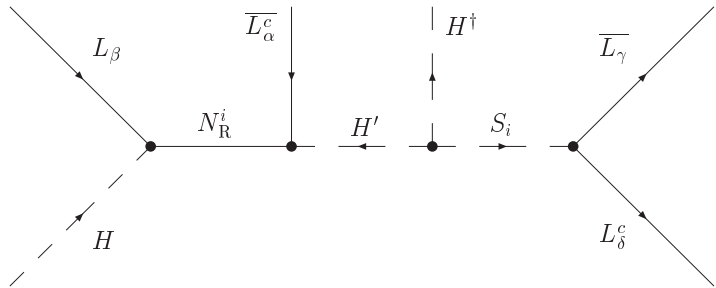}
 \caption{\label{fig:d=8case3}
Generation of dimension 8 operators which induce NSIs with matter after EW symmetry breaking in extensions of the SM with fermionic singlets $N_\mathrm{R}^i$ coupling to one lepton and one Higgs doublet
as well as singly charged scalars $S_i$ coupling to two lepton doublets (via topology 1 of figure \ref{fig:topologies}). 
}
\end{figure}

In the mixed case (c.f.\ diagram shown in figure \ref{fig:d=8case3}) with one coupling of a pair of lepton doublets to $S_i$ and one coupling of a lepton and a Higgs doublet to $N_\mathrm{R}^i$, constraints on the NSIs can be derived analogously to section \ref{sec:ConstraintsTypeB}. Furthermore, they are of comparable magnitude since the constraints on $v |Y_{\alpha i}/M_i| < v \sqrt{|c_{\alpha\alpha}^{d=6,kin}|}$ (c.f.\ Eqs.~(\ref{Eq:D=6MaboveEW}) and (\ref{Eq:D=6MbelowEW})) and 
$v |\lambda^i_{e \mu}/m_{S_i}|<  2.9  \cdot 10^{-2}$, 
$v |\lambda^i_{e \tau}/m_{S_i}| < 9.2  \cdot 10^{-2}$
(from the results of section \ref{App:AnitSymmDim6}) are of the same order. 
The associated dimension 8 operators are of the form
\begin{eqnarray}\label{Eq:Dim8CaseIII}
{\cal L}^{d=8}_{NSI,III} =  
c^{d=8,f,III}_{\alpha\beta\gamma\delta} 
(H^\dagger \bar L^c_\alpha) (L_\beta \cdot H) (\bar L_\gamma \cdot L^c_\delta)
\;.
\end{eqnarray} 
Again, we would like to note that the couplings required for generating the dimension 8 operators also give rise to dimension 6 operators which themselves produce non-standard interactions with matter as well as  non-standard interactions at the source and detector.

\section{Conclusions}

We have investigated how non-standard neutrino interactions (NSIs) with matter can be induced by new physics beyond the Standard Model (SM) and derived the corresponding constraints. One motivation for this study was that while NSIs at the source and detector are typically assumed to be strongly constrained, in many phenomenological studies very large NSIs with matter ($\varepsilon^{m,f}_{\alpha\beta}$ parameters ${\cal O}(1)$) are considered, saturating their weak direct bounds. 
To justify such large NSIs with matter (while escaping the stronger constraints that would stem from their charged fermion SU(2)$_\mathrm{L}$ doublet counterparts) it is often referred to specific classes of higher-dimensional operators. 
The goal of this study was to investigate the constraints on the NSI parameters if these operators are generated by explicit new physics beyond the SM.

In our analysis, we have focused on the tree-level realisations of dimension 6 and 8 operators which do not induce new interactions of four charged fermions (since these are already quite constrained). We have furthermore discarded the possibility of cancellations between diagrams with different messenger particles to circumvent constraints. The cases studied include the classes of effective higher-dimensional operators
mentioned above, which are often referred to as examples for ways to generate large NSIs with matter. 
A discussion of the restrictions and limitations of our analysis is given in section \ref{sec:restrictions}.

Regarding dimension 6 operators there are only two possibilities which satisfy the criteria defined in section \ref{sec:restrictions}: 
the anti-symmetric 4-lepton operator of Eq.~(\ref{Eq:AntisymmDim6}), generated from the exchange of virtual singly charged scalar fields $S^i$, and the dimension 6 operator of Eq.~(\ref{Eq:Dim6Kin}) modifying the neutrino kinetic terms, generated by the exchange of virtual fermionic singlets $N_\mathrm{R}^i$. The latter operator generates the NSIs in an indirect way, i.e.\ after canonical normalisation of the neutrino kinetic terms. 
For both possibilities we have derived improved bounds on the NSI parameters (c.f.\ table \ref{tab:boundsNSIMatterFromD=6} and table \ref{tab:boundsNSIMatterFromD=6Kin}). The bounds on the quantities  $|\tilde\varepsilon^{m}_{\alpha\beta}|$ are at least ${\cal O}(10^{-2})$.

We have then analysed the possibility to generate NSIs with matter from dimension 8 operators. 
Performing a systematic analysis of tree-level generations of operators with external fields 
$L,\bar L, f, \bar f, H, H^\dagger$, where $f$ can be $f_\mathrm{L}$ or $f^c_\mathrm{R}$   with $f_\mathrm{L} \in \{L_1,Q_1\}$ and $f^c_\mathrm{R} \in \{e^c_\mathrm{R},u^c_\mathrm{R},d^c_\mathrm{R}\}$,
we found that our criteria of section \ref{sec:restrictions} require that either lepton and Higgs doublets couple to SM singlet fermions (right-handed neutrinos) and/or that two lepton doublets couple to singly charged scalars fields $S^i$. 
Using the LEP bounds on the charged components of the additional scalar fields required for realising the dimension 8 operators and allowing for Yukawa couplings ${\cal O}(1)$, the NSI parameters    
$|\tilde\varepsilon^{m}_{\alpha\beta}|$ are constrained to be below ${\cal O}(10^{-2})$.

In summary, we have found that in the considered setup (c.f.\ section \ref{sec:restrictions}), NSIs with matter are considerably more constrained than assumed in many phenomenological studies, at least ${\cal O}(10^{-2})$.  
In some cases the bounds on the NSIs from sources which have previously been regarded as quite unconstrained have turned out to be even stronger than for conventional dimension 6 operators with structures $\bar L L \bar Q Q$ or $\bar L L \bar L L$ (which induce NSIs as well as interactions of four charged fermions). 
We have furthermore found that the generation of NSIs with matter always gives rise to additional NSIs at the source and/or detector of a possible future Neutrino Factory.   
These NSIs at the source and detector can, for instance, lead to ``zero distance'' neutrino flavour conversion effects which can be efficiently looked for in near detectors at future neutrino oscillation facilities.
While NSIs with matter with a strength below ${\cal O}(10^{-2})$ will be difficult to observe at currently planned or running experiments, they might be observed at envisioned Neutrino Factories or $\beta$-Beam facilities \cite{Bandyopadhyay:2007kx} and their possible impact on precision measurements of the neutrino parameters cannot yet be ignored. 
In order to determine the possible new physics effects in such high precision neutrino oscillation experiments, searches at near detectors in neutrino oscillation experiments, improved data from EW precision tests and rare lepton decays as well as the results from the LHC will play a crucial role.

\section*{Acknowledgements}
We would like to thank Carla Biggio, Mattias Blennow, Blanca Fern\'andez Mart\'{\i}nez, Belen Gavela and Miriam Tortola for useful discussions. This work was partially supported by The Cluster of Excellence for Fundamental Physics ``Origin and Structure of the Universe'' (Garching and Munich).

\providecommand{\bysame}{\leavevmode\hbox to3em{\hrulefill}\thinspace}

\end{document}